\documentclass[twocolumn,prb,showpacs,floatfix]{revtex4}
\usepackage{bm}
\usepackage{array}
\usepackage[dvips]{epsfig}
\usepackage{amsmath}
\usepackage{amsfonts}

\newcommand{\fitaa}{0.0846}
\newcommand{\fitbb}{4.51 \times 10^6}
\newcommand{\fitcc}{3.53}

\begin{document}
\author{B.A. DiDonna}
\affiliation{Department of Chemistry and Biochemistry, University of California Los Angeles,
Los Angeles, CA 90095, USA}
\author{D.C. Morse}
\affiliation{Department of Chemical Engineering and Material Science,
University of Minnesota, Minneapolis, MN 55455-0436, USA}

\title{Gellation of rigid filament networks}

\date{\today}

\begin{abstract}
We consider a model for gelation of rigid rods, in which rods that are
initially placed at random undergo diffusion, and form cross-links when they 
collide. In the limit of point-like cross-links, the number $N$ of croslinks 
per rod approaches $N \simeq \fitcc$. In a model with compliant cross-links 
of maximum length $\ell_c$, $N(t)$ increases with time as $N(t) \propto const. 
+ cL^{2}\ell_c \ln(t)$, where $c$ is concentration and $L$ is rod length.
\end{abstract}
\pacs{61.46.Df,82.70.Kj,87.15.Aa}
\maketitle

\section{\label{intro}Introduction}

The basic physics of networks of rigid and semiflexible filaments has proved 
to be important to many systems of current interest. Carbon nanotubes in a 
buffered solution have very strong localized interactions which can be modeled 
as permanent cross-links~\cite{hough.2006}. Similarly, microtubules and actin 
filaments in the cytoskeleton have long persistence lengths and are 
cross-linked by a number of chemical agents~\cite{alberts:88}. Still, there is 
very little knowledge concerning the generic properties of cross-linked gels 
created from long, thin filaments with high bending rigidity. There is not 
even a consensus on the expected connectivity of such gels. Without some
understanding of connectivity, it is impossible to formulate {\it ab initio} 
models for the electrical and thermal conductivity of these systems, much less 
their rheological properties.

%The connectivity of rigid filament networks it difficult 
%to model based on first principals. Unlike the 
%well studied case of flexible polymer gels~\cite{grest.1990},
%individual  polymer filaments are not in intimate 
%contact with there neighbors
%over most of their length. Instead, the number of contacts
%between filaments is highly topologically constrained. 
Previous numerical simulations of the gelation of rigid rod networks
\cite{balberg.1984,foygel.2005} have taken a purely geometrical 
constructive approach: In this aproach, straight filaments with a
nonzero diameter $d$ and a large aspect ratio $L/d$ are placed in a 
unit cell at random, and are assumed to be attached only if they 
overlap.  In a three dimensional system, the number of cross-links 
formed in this way depends directly upon the rod diameter, and 
vanishes in the limit of infinitely thin chains.  The situation is, 
of course, very different in a two dimensional model, for which this 
approach can yield a percolation threshhold for infinitely thin rods.
There is also a rich literature on both connectivity and rigidity 
percolation in networks of rods
~\cite{head.2003,wilhelm.2003,sahimi.1993,moukarzel.1999,jacobs.1997},
which rely on such static, constructive approaches to building a
network. 
However, when three dimensional networks are created in this way, they 
contain many near misses, which would be transformed into cross-links 
if the rods were allowed any mobility. The number of close approaches 
just outside the overlap cutoff scales with the density of the system.  
Thus, as we will show, this approach vastly underestimates the maximal 
number of connections that can be formed in a three dimensional system 
if the rods are allowed to diffuse. 

Here, we consider a simple dynamical model in which gelation instead
occurs as the result of Brownian motion of thin rods. Rods are initially 
placed at random and then undergo Brownian motion. Pairs of rods are 
assumed to irreversibly form cross-links whenever the distance of closest
approach falls below some ``capture radius".  Throughout this paper, we 
consider only ``flexible" cross-links that exert no torque, which thus 
do not introduce a constraint or bias on the angle between cross-linked 
pairs of rods.  The simplest variant of such a model is one in which 
rods cross-link when they collide, and form a permanent point-like 
(but rotationally flexible) cross-link at the point of collision. In 
this limit, the cross-linking would proceeds to a well defined saturation 
point: The creation of cross-links would continue until the system was
mechanically rigid, i.e., until the constraints imposed by the cross-links
allowed no further motion of the network. 

We may estimate a lower bound on the number of cross-links created in 
this idealized limit of point-like cross-links by simple degree of freedom 
counting: Each rod has 5 rigid motion degrees of freedom (three translations 
and two physically relevant rotations, excluding axial rotation). Each
cross-link introduces three constraints, or removes 3 degrees of freedom, 
by equating the positions of the points along two rods at which those 
rods collide. This argument thus predicts that the total ratio of cross-linkers 
to rods at the rigidity transition should be $5/3$. Since each cross-link 
connects two rods, the average number of rods to which a randomly chosen 
rod is connected is predicted to be $10/3 = 3.333$. 

This estimate is expected to underestimate the number of cross-links somewhat,
because it assumes that the constraints introduced by random cross-links are 
all linearly independent. It is possible for the system to form cross-links 
that are, in a sense, redundant. The simplest example of this occurs when a
a rod that is cross-linked at two points to a network that would remain rigid
if this rod was removed. The two cross-links required to immobilize this
rod introduce 6 constraints into the system of equations required to describe
the system, but remove only the 5 degrees of freedom associated with that 
rod, thus ``wasting" one constraint. As a result of this kind of redundancy, 
one might expect the average number of attachment points per rod to be 
slightly higher than $10/3$ (as is found to be the case). 

The idealized limit of point-like cross-links discussed above is, however,
very difficult to directly simulate, as well as being physically unrealistic
for many systems of interest.  The most straightforward way to simulate 
the Brownian motion of a partially gelled system of rods with point-like 
cross-links would be to introduce Langrange multiplier forces for each of 
the cross-links. A Brownian dynamics simulation of the system with such
cross-links would then require the solution of a large system of linear 
equations every time step in order to determine the constraint forces 
exerted by the cross-links. This would be prohibitively expensive for all
but very small systems or short simulations. 

Here, we instead consider a system of slightly compliant cross-links, in 
which the distance between the cross-link attachment points on a pair of 
rods is constrained to remain less than some maximum cross-link length 
$\ell_c$. A network with compliant cross-links retains some flexibility 
even when the number of cross-links would arrest all motion in a
corresponding network of point-like cross-links. As a result, the number
of cross-links in such a system never truly saturates, but increases more
and more slowly, as the addition of further cross-links requires more and
more rare thermal fluctuations. We find that this creates a very slow
logarithmic growth in the number of cross-links with time at late times.
We show, however, that it is possible to extract information about 
the number of cross-links that would be formed in the idealized limit 
of point-like cross-links by running simulations with several values of 
$\ell_c$ and extrapolating the results to $\ell_c = 0$. 

\begin{figure}

\centerline{\includegraphics[width=3in]{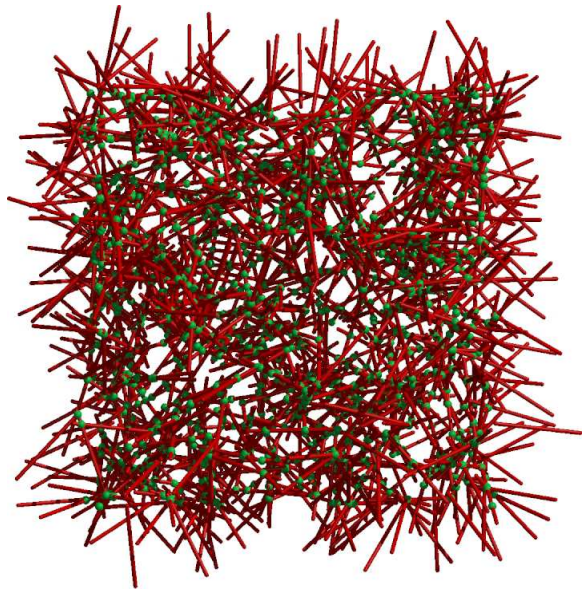}}

\caption{Random distribution of $1000$ rigid filaments
with density $\bar{c}=10$. 
For filaments which cross the boundaries, both periodic images are shown.
The diameter of the filaments are $0.0052$ times their length.}
\label{figure:image}
\end{figure}

%( Skip the following paragraph )

%We have found that for a random solution of rigid rods with 
%number concentration $c$ and length $L$, permanent cross-linking 
%will evolve at long times as
%$N\approx \fitcc + \fitaa c L^2 \ell_c 
%\log ( \fitbb (c L^3)^2 t)$, where $\ell_c$ 
%is the compliance length of the cross-links and $t$ is measured in 
%dimensionless simulation units. Substituting for the proper 
%units and constants associated with diffusion gives that
%one simulation time units is
%approximately $10^{-12} (\eta/\eta_{water})(L / 1 \ nm)^3$ seconds, 
%where $\eta_{water}$ is the viscosity of water.
%In the limit of zero diameter filaments and zero radius cross-links, there
%is no late time evolution in cross-link number. For this limit, the
%saturation cross-link number is a geometric constant independent 
%of filament concentration and slightly above the theoretical minimum
%connectivity required for rigidity.

\section{\label{sim}Simulation}

Our simulated systems consist of rigid filaments modeled by straight 
line segments of unit length $L=1$.  The initial state of the system 
is prepared by placing the rods in a simulational box with uniformly 
distributed random positions and orientations. We simulated dimensionless
rod number concentrations $\bar{c} \equiv cL^3$ in the range
$2<\bar{c}<2000$. Each system is run with at least $1000$ filaments and a 
box volume $\ge 10 L^3$. Periodic boundary conditions are enforced. 
The filaments are moved one at a time in random order via an unconstrained
Brownian dynamics algorithm.  For each filament the transverse, longitudinal, 
and rotational displacements were calculated separately, with drag
coefficients chosen so that the transverse and rotation 
diffusion constants were respectively
\begin{gather}
D_{\perp} = \frac{1}{2} D_{\parallel} \nonumber \\
D_{r} = 6 D_{\parallel} / L^2, \nonumber
\end{gather}
where $D_{\parallel}$ is the longitudinal diffusion constant~\cite{doi.edwards}.

In the simulations presented here, a permanent cross-link is created 
between two filaments whenever a Brownian dynamics move would have 
caused one of the filaments to cut through the other. 
Though not presented here, we found 
quantitatively similar results when the cross-linking criterion was 
that two filaments approached within a finite capture radius of one another.
The cross-link connects the two rods at their points of intersection. 
To inhibit bundling, only one cross-link was allowed between any two 
filaments. For dense systems and rigid filaments, bundling proved not 
to be an issue, since trapped entanglements did not allow filament 
alignment. 

The constraints of cross-link compliance 
and topology conservation are enforced via a Monte-Carlo acceptance 
criterion, as in Ref.~\cite{Shriram} -- motions which cause the 
center-lines of neighboring filament to cross or which cause the length
of any existing cross-link to exceed a maximum value are rejected. In 
some cases we also enforce a hard core potential between rods using the 
same technique. Unless otherwise stated, our simulated filaments have 
zero excluded volume.
Cross-links imposed no rotational constraints about their axes.

%Cross-links are formed between two fixed points on different filaments 
%according to a criterion evaluated at each time step. We simulated two 
%different criterion for cross-link formation. The first criterion was 
%based on a capture radius $r_c$: whenever the distance of closest
%approach between two filaments came within a predefined value $r_c$, a 
%cross-link was added between the points of closest approach on the 
%filaments.  The second criterion was collision, which is equivalent 
%to $r_c=0$: whenever a move would make two filaments cross each other, 
%that move is rejected, but the filaments are connected by a cross-link
%at their point of collision.  Once added, cross-links are permanent. 

%We found that the results for both forms of cross-linking 
%criterion were qualitatively similar. In the remainder of this paper, 
%we present data for the collision criterion, with $r_c=0$.

As we show in the next section, we can easily 
extrapolate from our data to the case of zero radius 
cross-links $\ell_c=0$. We note,
however, that direct simulation of cross-links that are completely
noncompliant to stretching is impracticable via any simulational technique.
We also comment that we are pursuing similar simulations using 
molecular dynamics on bead-chain polymers, but such systems are 
unable to approach the limit of infinite bend rigidity.

\section{\label{data}Results}

\subsection{Cross-link number}

\begin{figure}

\centerline{\includegraphics{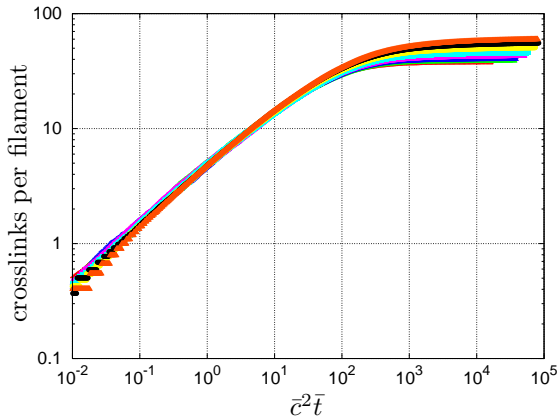}}

\caption{Time evolution of cross-linking number for $\bar{\ell}_c=0.0013$
and, from lowest to highest curves respectively,
$\bar{c}=100$, $150$, $200$, $300$, $400$, $600$, $800$, and $1000$.
The early time evolution of 
all the plots collapsed when ploted versus $\bar{c}^2 \bar{t}$, 
where time is measured in units of the 
rotational diffusion time $1/6D_r$.}
\label{figure:time}
\end{figure}

Immediately after
cross-linking was initiated the number of
cross-links was found to grow as 
$\bar{c} \sqrt{\bar{t}}$ (see Figure~\ref{figure:time}), 
where 
dimensionless time $\bar{t}$ is measured in units of the 
rotational diffusion time $1/6D_r$.
This power
law growth ends at a crossover time of order $\bar{t}
\sim 100/\bar{c}^{2}$. We associate this crossover time
with the time necessary for a filament to sense its
neighbors by diffusion, since the distance to nearest
neighbors (corresponding to the "cage diameter" in
the Doi model of entangled rigid rods) is proportional to
$1/\bar{c}$.
The power law growth was followed by a period of slow
logarithmic growth. The prefactor of the logarithm  
was observed to increase for
larger dimensionless cross-link radius $\bar{\ell}_c \equiv \ell_c/L$ or higher 
concentration $\bar{c}$. We simulated many combinations of 
these parameters in the range  
$0.0013 \le \bar{\ell}_c \le 0.013$ and $5 \le \bar{c} \le 2000$. 
All simulations were run
with dimensionless time-steps $10^{-9}$, because we found that the time
dependent evolution had nearly converged for time-steps on this 
order of magnitude.
Obtaining a half decade
of logarithmic growth on a system of $1000$ or more 
filaments typically took $12$ days on
a single 3.2 GHz Xeon based Linux system. Each data point 
presented here represents the average of four 
independent systems.

\begin{figure*}

\centerline{\includegraphics{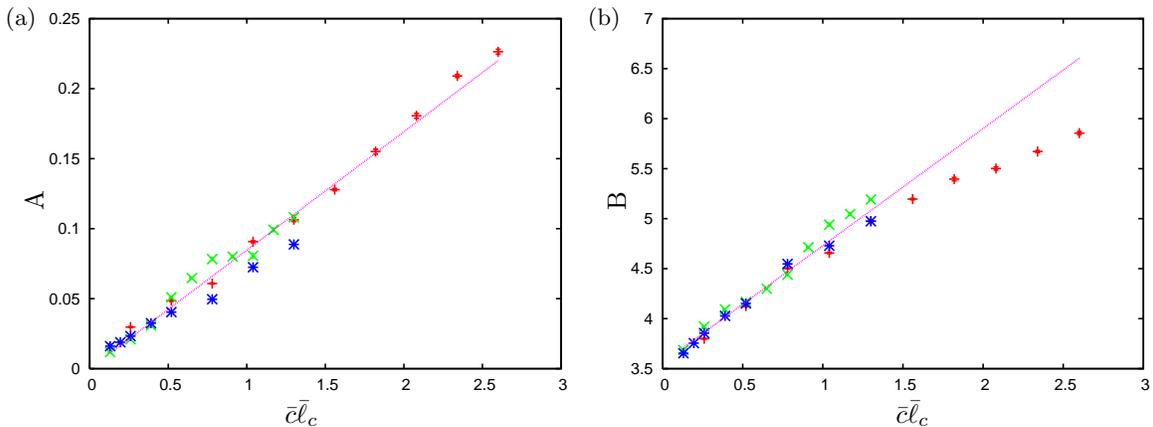}}

\caption{Fitting parameters $A$ and $B$ from a fit to the
form $A \log \left( \bar{c}^2 \bar{t} \right) +B$ for the late time evolution of
cross-link number, plotted as a function of the product $\bar{c} \bar{\ell}_c$. Crosses
are for runs with constant $\bar{\ell_c} = 0.0013$, stars are 
for runs with constant $\bar{c}=100$, and plus symbols are for runs with constant
$\bar{c}=200$. Straight lines 
are linear fits
of  $A$ to $\alpha \bar{c} \bar{\ell}_c$  and $B$ to $\beta+\beta' \bar{c} \bar{\ell}_c$}
\label{figure:sat}
\end{figure*}

Figure~\ref{figure:sat} shows our numerical fits
to the scaling data. We fit the late time
scaling in cross-link number for each concentration
to the form $A \log(\bar{c}^2 \bar{t}) + B$ via parameters $A$ and $B$.
Independent fits of $A$ and $B$ to powers of
$\bar{c}$ and $\bar{\ell}_c$ indicated that the overall scaling was 
consistent with a pure dependence on the dimensionless
combination $\bar{c} \bar{\ell}_c$. This simple dependence seemed 
more physically plausible, so we restricted our fits  
to this combination.
We fit the value of $A$ to a 
form $\alpha \bar{c} \bar{\ell}_c$ 
via parameter $\alpha$ and the coefficient $B$ to a form 
$\beta+\beta' \bar{c} \bar{\ell}_c$ via
the parameters $\beta$ and $\beta'$.  The parameter $\beta'$ may be 
absorbed into the logarithm as a multiplicative constant
$\gamma= \exp \left( \beta' / \alpha \right)$.

The result of this fitting procedure was the
following form for 
the cross-link number:
\begin{gather}
N = \beta + \alpha \bar{c} \bar{\ell}_c 
\log (\gamma \bar{c}^2 \bar{t}) \nonumber \\
\alpha = \fitaa \pm 0.0013 \nonumber \\
\beta = \fitcc \pm 0.033\nonumber \\
\gamma = \fitbb \pm 2.2 \times 10^5
\label{eq:satN}
\end{gather}
We chose to restrict our linear fit of the parameter $B$ to the
region $\bar{c}\bar{\ell}_c \le 1.5$, which seemed to 
converge linearly to the value $B=\fitcc$
for asymptotically small $\bar{c}\bar{\ell}_c$. Higher values measured 
for $\bar{c}=200$ and $0.078 \le \bar{\ell}_c \le 0.013$ 
deviated significantly from this linear fit. This deviation may result from 
three body interactions for very large values of cross-link radius 
$\bar{\ell}_c$, or it may simply result from our time-step being to large 
under these very extreme conditions. Unfortunately, since these 
simulations were run at the limit of our current 
computational capability, clarification will have to wait for future 
work. Though data is not included here, 
the cross-link number was seen to
deviate upward from the fitted linear relation for $\bar{c} \le 10$ --
we will discuss this discrepency further below. 

In the next section we develop arguments, based on additional
measurements, for the physical origin of each term in 
Eq.~\ref{eq:satN}. We link the constant term to the 
rigidification threshold linking number for unstretchable cross-links 
and the term linear in $\bar{c}$ and $\bar{\ell}_c$ to localized motion
within the cross-link length. We attribute the
logarithmic growth term to collective motion of the system. The
logarithm reflects slow, glassy dynamics. 

\subsection{Cross-link statistics}

To shed light on the origins of the cross-linking number
in Eq~\ref{eq:satN}, we examine another informative statistic:
the radial distribution function of unbound neighbors to a
given rod. It can be shown~\cite{shriram.thesis} that for a random
spatial distribution of 
non-interacting rods, given a test rod, the probability that another 
rod will have a distance of closest approach between a radius 
$r$ and $r + dr$ is given by
\begin{equation}
P(r)dr = \frac{\pi}{2} c L^2 dr.
\label{eq:rdf}
\end{equation}
Thus, without cross-linking,
the probability of finding a neighboring rod with closest 
approach at any given radius is independent of the radius.

\begin{figure}

\centerline{\includegraphics{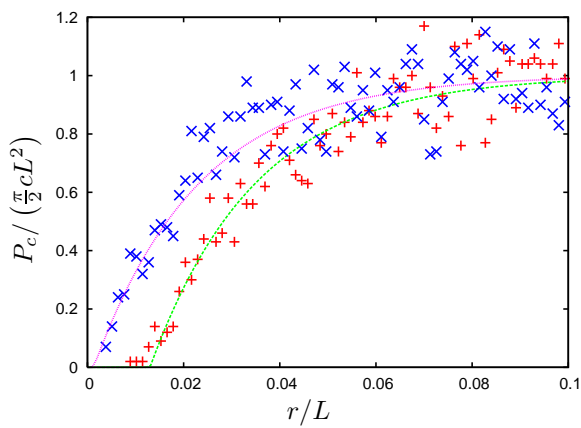}}

\caption{Radial distribution function for distance of
closest approach of unlinked neighbor filaments. All data is 
for a concentration of $cL^3=100$. The upper and lower
data sets are respectively for $\ell_c/L =0.0013$ and 
$\ell_c/L =0.013$. Solid lines are the predicted 
distribution from Eq~\ref{eq:rdfc} with $b=1$.}
\label{figure:rdf}
\end{figure}

Figure~\ref{figure:rdf} shows the measured distribution 
for the distance of closest approach of unlinked 
neighbor filaments to a given test filament, averaged over all
filaments. The measured probability function 
is well fit by the function
\begin{equation}
P_c(r) \approx
\begin{cases}
0  & r \le b \ell_c \\
 \frac{\pi}{2} c L^2
\left(
1-\exp \left(- \frac{\pi c L^2}{2 \times \fitcc} 
\left( r - b \ell_c \right)\right)
\right) & r > b \ell_c 
\end{cases}
\label{eq:rdfc}
\end{equation}
where the value of $b$ grows with time. At any given time,
when $P(r) - P_c(r)$
is integrated from zero to infinity it gives exactly 
Eq.~\ref{eq:satN} for the number of cross-links which each filament has
captured on average. More spefically, the constant term in 
Eq.~\ref{eq:satN} comes from integrating the exponential term
in Eq~\ref{eq:rdfc}, and
all the other terms
in Eq.~\ref{eq:satN} comes from itegrating the 
depleted region between $0 < r < b \ell_c$ in 
Eq~\ref{eq:rdfc}. 

We interpret Eq~\ref{eq:rdfc} as proof that 
the extrapolated linkage 
number $N=\fitcc$ is exactly the linkage number
required for rigidification of line segments in
$3$-dimensional space, which we shall call $N_r$. 
If the cross-links had no stretching compliance
($\ell_c=0$), then after $N_r$ cross-links per filament had been added
the lattice would become completely elastically rigid. In this case, there would
be no further spatial fluctuations, and the cross-linking would cease.
Prior to the rigidity transition, the 
filaments should diffusively explore their
local environment up to the point when they have
encountered and linked to $N_r$ 
other rods on average. Since the probability of finding a rod at any 
given radius is a flat distribution, the probability of 
finding $N_r$ rods {\it within} radius $r$ is exponential, with form
$$
\exp \left(- \frac{\pi c L^2}{2 N_r} r\right).
$$
This form correspond exactly to the exponential in Eq~\ref{eq:rdfc}
if we make the identity $N_r=\fitcc$. This measured value is close
to the theoretical lower bound of $N_r=10/3$ which we derived in the 
introduction.

We note that as $\bar{c}$ approaches the value
$\frac{2}{\pi} \times \fitcc \sim 2.2$, the decay length in Eq.~\ref{eq:rdfc}
approaches $L$, so that rod end effects will become more important. 
In this limit the rods will interact more like point-like particles. 
This is consistent with our observation (not shown) of a deviation upward from
the logarithmic  fit in Eq.~\ref{eq:satN} for $\bar{c} \le 10$.

The effect of finite cross-link length is less straightforward to interpret. 
After the rigidity connectivity $N_r$ has been reached the only
possible fluctuation motion of any filament is within the 
stretch compliance of cross-links to its neighbors. The 
density of neighbors within the cross-link length is higher 
than outside, and without collective motion of its neighbors,
there is very little room for a filament to move around in. 
Our results suggest that filaments quickly explore a
region of radius $r_c$, producing the linear term in Eq~\ref{eq:satN}.
At long times, they continue to sample and link to neighbors in the
space around them, increasing their sampling radius at a logarithmic
rate. Furthermore, the exponential in Eq~\ref{eq:rdf}
is shifted by this sampling radius,
so that $P_c(r)$ goes smoothly to zero at the edge of the
depletion zone. This is consistent with the slow, glassy dynamics
of large scale collective motions in the presence of deep potential 
wells.

%\begin{figure}

%\centerline{\includegraphics{figure5.eps}}

%\caption{Distribution function for the length of cross-links. 
%The distribution is normalized to pass through $1$ at $\ell=0$.}
%\label{figure:cdf}
%\end{figure}

%In Fig.~\ref{figure:cdf}, we plot a complimentary statistic to 
%that of Fig.~\ref{figure:rdf}:
% the time-average distribution of distances 
% between cross-linked points. 
% For early time or low density
%solutions, the distribution of cross-link separations is quadratic,
%$P_l(\ell) \propto \ell^2$,
%which is consistent with completely unconstrained motion within
%the spherical region defined by $\ell \le \ell_c$. More
%interesting behavior is found for late time systems at 
%high densities, where the cross-linking number is above the
%minimal rigidity number of $N_r=\fitcc$. For this case, we find
%that 
%$$
%P_l(\ell) \propto \ell^2 \exp \left( \mu \left(\frac{\ell}{\ell_c}\right)^2\right),
%$$
%where $\mu$ grows up to a value of $\mu \approx 0.67$ and then seems 
%to level off (within experimentally observable times).
%This dependence suggests an entropic potential pulling
%on the cross-links. At this time we cannot explain the measured
%value of stiffness coefficent $\mu$. The observation of this 
%stiffness is consistent with our observation of 
%slow cross-linking dynamics for dense systems, as discussed above. 
%The filaments are free to explore the region within the lengths of
%their cross-links, but there is significant entropic damping of
%this motion for dense systems.

\begin{figure}

\centerline{\includegraphics{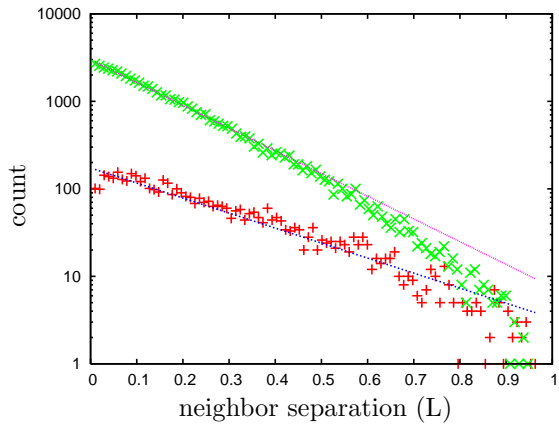}}

\caption{Distribution of cross-link separation lengths for late time 
systems with $\bar{\ell}_c=0.0013$ and $\bar{c}=150$ (lower curve) or
$\bar{c}=1000$ (upper curve). Straight lines are theoretically 
predicted exponential 
curve for random cross-link placement, disregarding end effects. }
\label{figure:clinksep}
\end{figure}

\begin{figure}

\centerline{\includegraphics{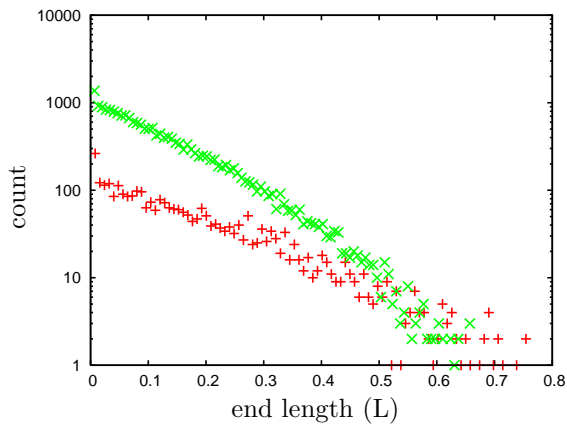}}

\caption{Distribution of lengths between filament ends
and nearest cross-link for late time 
systems with $\bar{\ell}_c=0.0013$ and $\bar{c}=150$ (lower curve) or
$\bar{c}=1000$ (upper curve). }
\label{figure:clinkends}
\end{figure}

Finally, we analyze the statistics of cross-linking. 
In Figure~\ref{figure:clinksep}, we plot the distribution of separations 
between neighboring cross-links. As expected, the distribution is exponential, 
with decay constant equal to the average density of cross-links per unit length.
The exponential decay is modified slightly by finite filament length effects at
separations approaching $L$. In Figure~\ref{figure:clinkends}, we plot the 
distribution of lengths
of ``dangling" ends beyond the last cross-link on each filament, which 
is also exponential.
These observed distributions are consistent with spatially
random placements of cross-links along the filament, and are qualitatively
the same as for flexible polymers.~\cite{grest.1990}

\section{\label{sec:discussion}Discussion}

We have found a new generic number for 
the maximum cross-linking in the limit 
of vanishing cross-link length, $N_r=\fitcc$. We
further speculate that this is the as yet undiscovered 
rigidification threshold coordination number for a random lattice of
rigid, line-like filaments in $3$ dimensions. We have also
found that the cross-link number depends logarithmically on time and 
linearly on both cross-link length and filament concentration, and we have
measured the linear coefficient.

It is worth comparing our results to percolation based 
Monte-Carlo studies without particle 
dynamics. Foygel et al.~\cite{foygel.2005} found that the
percolation threshold for randomly 
distributed but stationary rods with aspect
ratio $a$ occured at number density 
$\bar{c} \approx 0.76 a$ with an average cross-linking number
$N=1.2$. Comparing their
simulations to ours, their aspect
ratio is equivalent to $1/(2 \bar{\ell}_c)$. Thus for $\bar{\ell}_c=0.0013$, a typical value in 
our simulations, {\it connectivity} percolation would not occur 
below $\bar{c} \approx 290$, and even at this value filaments would have
around $1.2$ overlaps on average. Thus, the addition of dynamics
and localized bonding between filaments dramatically increases
the connectivity of the filament suspension.

An future extension to this work will be the addition 
of a finite filament bend modulus. For large bend modulus, we expect the transverse
filament fluctuations will be equivalent to an increase in capture radius. This
research is under way. It would also be interesting to study the effect of the
initial configuration on the cross-link number -- initial alignment of the
filaments might greatly increase the tendency towards bundling, thereby increasing the
cross-link number significantly.

\section*{Acknowledgments}

BD thanks Gary Grest and Pieter In't Velt for enlightening discussions.
BD acknowledges partial
support from the Institute for Mathematics and its Applications with funds provided by the
National Science Foundation.

\end{document}